\documentclass[fleqn,twoside]{article}
\usepackage{espcrc2}

\usepackage{graphicx}
\usepackage[figuresright]{rotating}

\def\amu{a_\mu}
\def\amuh{a_\mu^{{\mathrm{had}}}}
\def\MZ{M_Z}

\def\dah{\Delta\alpha^{(5)}_{\rm had}}
\def\dahs{\Delta\alpha^{(5)}_{\rm had}(s)}
\def\dahz{\Delta\alpha^{(5)}_{\rm had}(\MZ^2)}
\def\dah0{\Delta\alpha^{(5)}_{\rm had}(-s_0)}

\def\PLB{{\em Phys. Lett.}  B }

\newcommand{\be}{\begin{equation}}
\newcommand{\ee}{\end{equation}}
\newcommand{\ba}{\begin{eqnarray}}
\newcommand{\ea}{\end{eqnarray}}
\newcommand{\bea}{\begin{eqnarray*}}
\newcommand{\eea}{\end{eqnarray*}}
\newcommand{\bet}{\begin{center} \begin{tabular}}
\newcommand{\ent}{\end{tabular} \end{center}}

\newcommand{\mz}{M^2_Z}

\newcommand{\al}{\alpha}

\newcommand{\sigha}{\sigma_{\rm had}}

\newcommand{\bary}{\begin{array}}
\newcommand{\eary}{\end{array}}
\newcommand{\noi}{\noindent}

\newcommand{\nn}{\nonumber}
\newcommand{\gv}{\mbox{GeV}}
\newcommand{\mv}{\mbox{MeV}}

\newcommand{\bit}{\begin{itemize}}
\newcommand{\eit}{\end{itemize}}

\newcommand{\dalh}{\Delta \alpha^{\rm had}}
\newcommand{\epm}{e^+e^-}
\newcommand{\ra}{\rightarrow}

\newcommand{\AmS}{{\protect\the\textfont2
  A\kern-.1667em\lower.5ex\hbox{M}\kern-.125emS}}
\newcommand{\sinf}{\sin^2 \Theta_f}
\newcommand{\sini}{\sin^2 \Theta_i}
\newcommand{\cosi}{\cos^2 \Theta_i}
\newcommand{\sing}{\sin^2 \Theta_g}
\newcommand{\dal}{\Delta \alpha}
\newcommand{\sinW}{\sin^2 \Theta_W}
\newcommand{\Gmu}{G_\mu}
\newcommand{\dro}{\Delta \rho}
\newcommand{\ep}{\;\:.}
\newcommand{\myit}{\noi ~$\bullet$ }

\title{The role of $\sigma(\epm \to {\rm hadrons})$ in precision tests of the Standard Model}

\author{%
F.~Jegerlehner\address{Deutsches Elektronen-Synchrotron (DESY), Platanenallee 6,
                       D-15738, Zeuthen, Germany}%
               \thanks{Work supported in part by
                       TMR, EC-Contract No.~HPRN-CT-2002-00311 (EURIDICE)}
}
\begin{document}
\onecolumn{
\renewcommand{\thefootnote}{\fnsymbol{footnote}}
\setlength{\baselineskip}{0.52cm}
\thispagestyle{empty}
\begin{flushright}
DESY 03--189 \\
November 2003\\
\end{flushright}

\setcounter{page}{0}

\mbox{}
\vspace*{\fill}
\begin{center}
{\Large\bf The role of $\sigma(\epm \to {\rm hadrons})$ in \\ precision
tests of the Standard Model}

\vspace{5em}
\large
F. Jegerlehner\footnote[4]{\noindent Invited talk at SIGHAD03: 
Worskhop on Hadronic Cross Section at Low Energy,
Pisa, Italy, October 8-10, 2003. Work supported in part by 
TMR, EC-Contract No.~HPRN-CT-2002-00311 (EURIDICE)} 
\\
\vspace{5em}
\normalsize
{\it   DESY Zeuthen}\\
{\it   Platanenallee 6, D--15738 Zeuthen, Germany}\\
\end{center}
\vspace*{\fill}}
\newpage

\begin{abstract}
I present a summary of recent developments as they have been discussed
in the many interesting contributions to this workshop (SIGHAD03 at
Pisa).
\vspace{1pc}
\end{abstract}

\maketitle

\section{$\alpha(M_Z)$ IN PRECISION PHYSICS}
Hadronic cross section measurements in $\epm$--annihilation are an
indispensable input for the estimation of the non--perturbative
hadronic contribution the shift in the fine structure constant,
one of the fundamental input parameters for making precise theoretical
predictions for electroweak processes in standard electroweak theory
(SM). The errors of the experimental data directly affect and limit
the precision at which theoretical predictions can be made.  Thus
the uncertainties of the hadronic contributions to the effective fine
structure constant $\alpha(E)$ ($E$ the energy scale) are a serious
problem in electroweak precision physics~\cite{LEP}. At present the
most precisely known parameters are $\alpha$, $G_\mu$ and $M_Z$ and
typical (pseudo-) observables for which precise predictions are
possible are $\sinf ,~ v_f,~ a_f$ defined via the $Z\bar{f}f$
neutral current vertex ($f$ any lepton or quark) or the gauge boson
parameters $M_W,~
\Gamma_Z,~ \Gamma_W,~ \cdots$. However, it is not the low energy fine
structure constant $\alpha$ itself, but the effective value
$\alpha(M_Z)$ which is the appropriate input parameter for predicting
properties of the weak gauge bosons $Z$ and $W$. The two are related
by non--perturbative physics, to be discussed below. Present
accuracies of the basic parameters are the following: $\frac{\delta
\alpha}{\alpha}
\sim 3.6 \times 10^{-9}$, $\frac{\delta G_\mu}{G_\mu} \sim 8.6 \times
10^{-6}$, $\frac{\delta M_Z}{M_Z} \sim 2.4 \times 10^{-5}$,
$\frac{\delta \alpha(M_Z)}{\alpha(M_Z)} \sim 1.6 \div 6.8 \times
10^{-4}$ (present), while $\frac{\delta \alpha(M_Z)}{\alpha(M_Z)} \sim
5.3 \times 10^{-5}$ is what will be required for precision physics at
a future linear collider (like TESLA). Note the dramatic loss in
acuracy by five orders of magnitude in going from $\alpha=\alpha(0)$
to $\alpha(M_Z)$. A typical uncertainty in $\alpha$ is $\delta \Delta
\alpha(M_Z)=0.00036$ which contibutes to the uncertainty $\delta
\sin^2 \Theta_{\rm eff}=0.00013$, which affects substantially the
Higgs mass upper bound which derives to a large extent from the
$\sin^2 \Theta^\ell_{\rm eff}$ ($\ell=e,\mu,\tau$) measurements. The
LEP/SLD result at present is $\sin^2
\Theta^\ell_{\rm eff}=\frac14 (1-\frac{v_\ell}{a_\ell}) =0.23148 \pm
0.00017$.   Most precision observables
also receive perturbative QCD (pQCD) contributions, the accurate
evaluation of which is depending on a precise knowledge of the QCD
parameters $\alpha_s,\; m_c, \; m_b,\; m_t$. Corresponding efforts in
pQCD, lattice QCD, chiral perturbation theory and/or exploiting sum
rules are mandatory in this respect. An important set of quantities
which directly depend on $\dal$ are the mixing parameters $\sini$
obtained from $\al$, $\Gmu$ and $M_Z\;$ via the generalized Sirlin
relation: 
\small
\ba
\sqrt{2}\:G_\mu\:M_Z^2\:\sini\: \cosi =
\frac{\pi \al}{1-{\Delta r_i}} =
\frac{\pi \al(M_Z)}{1-{\Delta \bar{r}_i}}\!\!\!
\ea 
\normalsize
with effective fine structure constant
\be
\alpha(M)=\frac{\al}{1-{\Delta \alpha(M)}}\;\;.
\ee
Various definitions of $\sini$, which
coincide at tree level (i.e., for ${\Delta r_i}\sim 0$) are the
following: from weak gauge boson masses, from electroweak gauge
couplings and from the neutral current couplings of the charged
fermions one defines $\sinW = 1-\frac{M_W^2}{M_Z^2}$, $\sing =
e^2/g^2$ or $\sinf =\frac{1}{4|Q_f|}\;\left(1-\frac{v_f}{a_f} \right)$
($f\neq \nu $), respectively. The radiative correction $\Delta r_i
=\Delta r_i( \al ,\: \Gmu ,\: M_Z ,\:m_H,
\: m_{f\neq t},\:m_t)$ distinguishing
the different definitions depend on all SM parameters and thereby
provide an indirect constraint on $m_t$, $m_H$ or possible new
physics, which also would contribute. The leading corrections $\dal$
and $\dro$ (leading $m_t$-dependence) enter in the form $\Delta r_i=
\dal + \Delta \bar{r}_i=
\dal - f_i(\sini)\:\dro + \Delta r_{i\:\mathrm{rem}}$, exhibiting an
universal dependence on $\dal$. The predictions for $M_W$, $A_{LR}$,
$A^f_{FB}$, $\Gamma_f$, $\cdots$ are thus directly affected by the
error $\delta \Delta \alpha$ which translates into uncertainties
$\delta M_W$, $\delta \sini$, etc.\footnote{see plot
files: {\tt w03\_sef2\_theta.eps, w03\_mw.eps} at 
\hfill \\ \hspace*{4mm} {\tt \footnotesize
lepewwg.web.cern.ch/LEPEWWG/plots/winter2003/}}. The indirect Higgs
boson mass ``measurement'' yields $m_H= 96^{+60}_{-30}$ GeV
($\epm$--data based evaluation of $\dal$)\footnote{If one would base
the evaluation on $\tau$-data instead, the central value would be
shifted by about $\delta m_H \sim - 10 $ GeV.}.
For further details I refer to~\cite{LEP,Zeuthen2003,Pietrzyk}.

\section{RECENT EXPERIMENTAL AND \hfill \\ THEORY INPUT}
Recent advances/issues in the evaluation of the hadronic vacuum
polarization (VP) are based on the following results:

\myit Updated results from the 
precise measurements of the processes $e^+e^- \to \rho \to
\pi^+\pi^-$, $e^+e^- \to \omega \to
\pi^+\pi^-\pi^0$ and $e^+e^- \to \phi \to K_LK_S$ performed by the
CMD-2 collaboration have appeared
recently~\cite{CMD2,Fedotovitch}. The update was necessary due to an
overestimate (partally missing subtraction of VP
effect in the Bhabha- and $\mu\mu$-channels) of the integrated
luminosity in the previous analysis which was published in
2002~\cite{CMD}. A more progressive error estimate (improving on
radiative corrections, in particular) allowed a reduction of the
systematic error from 1.4\% to 0.6 \%. Subtracting the vacuum
polarization systematically at time-like scales leads to
lower cross-sections. Some other CMD-2 and SND data at energies
$E< 1.4$ GeV have become available as well. CMD-2 now is in good
agreement with SND results on the 4$\pi$ channels.

\myit Before in 2001 BES-II published their final 
$R$--data which, in the region 2.0 GeV to 5.0 GeV, allowed to reduce
the previously huge systematic errors of about 20\% to 7\%
~\cite{BES,Hu}.

\myit After 1997 precise $\tau$--spectral functions
became available~\cite{ALEPH,OPAL,CLEO,Davier1} which, to the extent
that flavor $SU(2)$ in the light hadron sector is a good symmetry, allows
to obtain the iso--vector part of the $\epm$--cross
section~\cite{tsai}. This possibility has first been
exploited in the present context in~\cite{ADH98}. First results on
$\tau$-spectra from BELLE have been presented at this
meeting~\cite{Hayashii}.

\myit With increasing precision of the low energy
data it more and more turned out that we are confronted with a serious
obstacle to further progress: in the region just above the
$\omega$--resonance, the iso-spin rotated $\tau$--data, after being
corrected for the known iso-spin violating effects, do not agree with
the $\epm$--data at the 10\% level~\cite{DEHZ}. Before the origin of
this discrepancy is found it will be hard to make further progress in
pinning down theoretical uncertainties.

\myit In this context iso-spin breaking effects in the 
relationship between the $\tau$-- and the $\epm$--data have been
extensively investigated in~\cite{CEN,Ecker}. The question remains whether
all possible iso-spin violating effects have been taken into account
in which case the discrepancy would have to be attributed to
experimental problems.

\myit A new bound $\delta a_\mu(0.6-2.0 \gv)<0.7\:
\times 10^{-10}$~\cite{Eidelman03} for the contributions of 
$\pi\pi\gamma,\pi\eta\gamma$ which include decay products from
$\pi^0\gamma,\sigma\gamma,f\gamma,a_1\gamma$ yields a severe
constraint on possible missing contributions reported
elsewhere~\cite{Narison:2003ur,Dubnicka}.\\
{\bf Ongoing activities:}\\
{\bf 1) $R$ from radiative return at meson factories:}

\myit New results for hadronic $\epm$ cross--sections
are expected soon from KLOE, BABAR and BELLE. These experiments,
running at fixed energies, are able to perform measurements via the
radiative return
method~\cite{RR,Khoze:2002ix,KLOE,Valeriani,Davier2,Eidelman2}. Preliminary
results presented recently by KLOE agree very well with the final
CMD-2 $\epm$--data~\cite{Valeriani}.  KLOE is close to
finalizing results for the $\pi\pi$ channel.  BABAR has presented
interesting preliminary results, in particular on the 4$\pi$
channels~\cite{Davier2}. The feasibility of such cross-section
measurements has also been studied at BELLE~\cite{Eidelman2}.\\
{\bf 2) $R$ from energy scans:}

\myit $R$ measurements at CLEO~\cite{Dytman} will resolve 
the Mark I vs. CB ``discrepancy''. Measurements at the energies 7.0,
7.4, 8.4, 9.4 10.0 and 10.3 GeV will be performed.\\ 
{\bf Future plans:}

\myit An upgrade of BEPCI/BESII is in progress: 
BEPCII/BESIII will run at energies 2.2, 2.6 and 3.0 GeV and measure
$R$ at precisions 5.5, 3.4 and 3.4 \% ([7.6, 7.0 and 5.6] \% now),
respectively~\cite{Hu}.

\myit The upgrade of VEPP-2M to VEPP-2000 is in 
progress and will allow to measure $R$ in the energy range 0.4-2.0 GeV
in the years 2005-2010.  The luminosity will increase by a factor of
10, the detectors CMD-2 and SND should gain a factor of 2 in
precision~\cite{Eidelman1}. This will provide important improvements 
for the presently most problemetic energy range 1.4 - 2 GeV.

\myit At DA$\Phi$NE an upgrade to 2 GeV for a $R$-scan 
is under discussion~\cite{Denig}.\\ 
{\bf Theory issues:}

\myit Last but not least an important change in the hadronic
contribution to $\amu$ was the change in sign of the leading hadronic
light--by--light contribution ($\pi^0$
exchange)~\cite{Nyffeler,Knecht:2001qf,Bijnens:1995xf,Hayakawa:1997rq}.

\myit Progress was made also in calculating the radiative 
corrections to $\pi^+\pi^-$ production for energy scans, for inclusive
measurements in radiative return~\cite{Hoefer:2001mx,Arbuzov:2003xf} and in photon
tagging~\cite{RR,Khoze:2002ix,Rodrigo} relevant at the meson ($\Phi$,$B$)
factories.

Some of these results have substantially influenced the precision of
the evaluations of the hadronic VP contribution to
$\alpha_{\mbox{\footnotesize QED}}(M_Z)$ and $(g-2)_\mu$ since
1995~\cite{EJ95}. The present status is reviewed in the following.
All of the forthcoming measurements are able to substantially improve
future evaluations.

{\bf Note concerning future developments:} In order to get $\sigha$ at
a precision better than 1\% in regions where data must be used in the
dispersion integrals, the following aspects should be keep in
mind. Also in future two different methods for measuring low energy
hadronic cross sections will be applied: the enegy scan and the
radiative return or photon tagging method. Both methods have their own
advadtages and problems and require appropriate efforts on the theory
side. The photon tagging method is one power in $\alpha$ higher order
and requires a correspondingly higher effort on the theory side. The
lower cross section is compensated for by the dramatically enhanced
cross section at the resonsnce at which the maschine is running. Cross
checks of results from both methods are important. Specific problems
are encountered at very low energies, where the total cross section
only can be obtained by adding up exclusive channel measurements
(particle identification etc.). In a transition region at medium
energies one urgently requires cross checking of exclusive
vs. inclusive measurements. At higher energy finally inclusive
measurements are done and simplify live considerably.  Theory has to
improve on radiative corrections for Bhabha scattering (small/wide
angle) as a reference process for $\mu^+\mu^-$ for normalization/cross
check/event separation and finally the hadron--production
$\pi^+\pi^-,\: \cdots$ itself. Important questions concern: How to
estimate the error from missing higher order? How to estimate model
dependence of final state radiation (FSR) from hadrons?  Do we have
background under control?  e.g.: $e^+e^- \to e^+e^- \pi^+
\pi^-$ in inclusive measurement of $\pi^+ \pi^-$ is at 1-2\% 
level\footnote{In Ref.~\cite{Hoefer:2001mx} the high energy
approximation has been considered which overestimates the effect, as
has been pointed out in~\cite{Czyz:2003gb}, recently.} at low
invariant $M^2_{\pi\pi}$. Let me point out that the presently aimed
precision requires to check the proper treatment of VP subtraction: we
now need to subtract the true time--like $\alpha(s)$ in
s--channels. In Bhabha scattering programs one has to carefully
distinguish t--channel ($\alpha(t)$) and s--channel ($\alpha(s)$)
subtractions. One lesson we have learned in particular from precision
physics at LEP/SLC: we in any case need dedicated comparisons of
different calculations/generators, in order to make sure that things
are under control. See also the
contributions~\cite{Rodrigo,Sibidanov,CaCa}. A serious problem remains
the radiation of photons by the final state hadrons, which we cannot
calculate reliably. Models like scalar QED for the photon radiation by
pions decsribe well soft real and virtual photons but fail at some
point to model hard real and virtual photons. The problem is
particularly serious for charged current processes like $\tau$--decay
(see below) where large UV logs come into play. For steps towards a
systematic treatment within the effective Lagrangian approach
see~\cite{CEN,Ecker,Scimemi}.

More details on theoretical aspects of radiative return 
measurements of $R$ are discussed in~\cite{Kuhn}. 
For more information on the status and prospects of $R$ 
measurements see~\cite{Zhao}.

\section{EVALUATION OF $\alpha(M_Z)$}
The non-perturbative hadronic contribution $\dahs$ to the photon
vacuum polarization $\dal =\Pi'_\gamma(0)-\Pi'_\gamma(s)$ can be
evaluated in terms of $\sigma(e^+e^- \to {\rm hadrons})$ data via a
dispersion relation\\[-12mm]

\centerline{%
\includegraphics[scale=0.85]{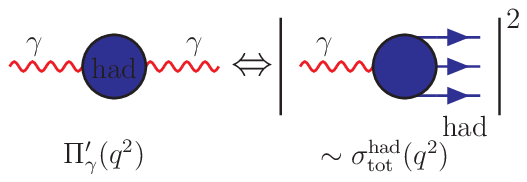}}

\vspace*{-5mm}

\noi which may be written as
\ba
\dahs &=& - \frac{\alpha s}{3\pi}\;\bigg(\;\;\;
{\rm \footnotesize P}\!\!\!\!\!\!\!\!
\int\limits_{4m_\pi^2}^{E^2_{\rm cut}} ds'
\frac{{ R^{\mathrm{data}}_\gamma(s')}}{s'(s'-s)}
\nn \\ &&~~~+ {\rm \footnotesize P}\!\!\!\!\!\!\!\!
\int\limits_{E^2_{\rm cut}}^\infty ds'
\frac{{ R^{\mathrm{pQCD}}_\gamma(s')}}{s'(s'-s)}\,\, 
\bigg)
\ea
with $R_\gamma(s) \equiv 
\sigma^{(0)}(e^+e^- \rightarrow \gamma^*
\rightarrow {\rm hadrons})/ \frac{4\pi \alpha^2}{3s} 
$.
My update for $M_Z=$ 91.19 GeV utilizing
$R(s)$ data up to $\sqrt{s}=E_{cut}=5$ GeV
and for the $\Upsilon$ resonance--region between 9.6 and 13 GeV
and perturbative QCD from 5.0 to 9.6 GeV
and for the high energy tail above 13 GeV yields
\ba
\Delta \al _{\rm hadrons}^{(5)}(\mz)   &=&  0.027690 \pm 0.000353 \\
\alpha^{-1}(\mz) &=& 128.936 \pm 0.048 \ep \nn
\ea
With more theory input, based on the Adler function method (see
below), we obtain (see Fig.~\ref{fig:aldist})
\ba
\label{eq:adler}
\Delta \al _{\rm hadrons}^{(5)}(\mz)   &=&  0.027651 \pm 0.000173 \\
\alpha^{-1}(\mz) &=& 128.939 \pm 0.024 \ep \nn
\ea

\vspace*{-5mm}

\begin{figure}[ht]
\begin{picture}(120,60)(15,27)
\includegraphics[scale=0.45]{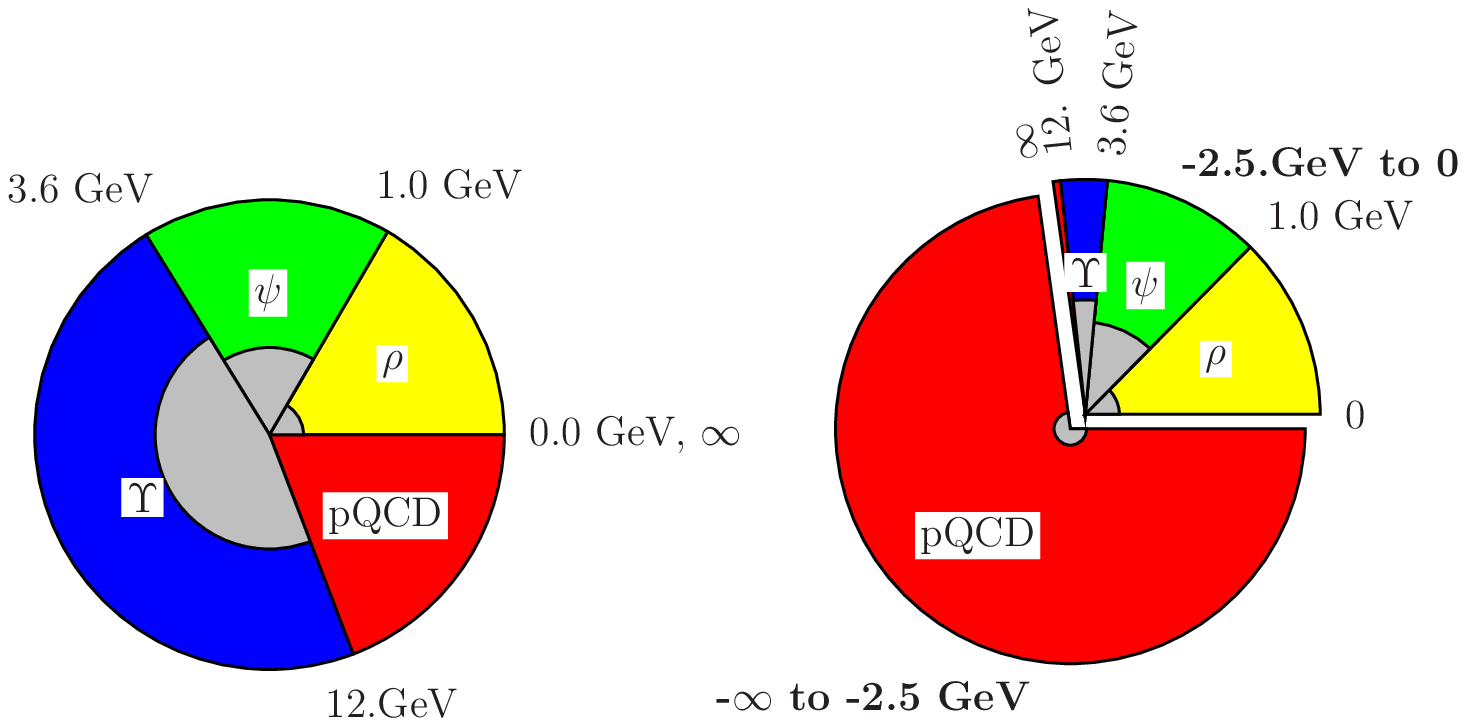}
\end{picture}

\vspace*{-2mm}

\caption{Comparison of the distribution of contributions and errors
(shaded areas scaled up by 10) in the standard (left) and the Adler
function based approach (right), respectively.}
\label{fig:aldist}


\end{figure}

Other recent evaluations agree well with these results, although the
methods are quite different in spirit:\\[-2mm]
 
\begin{tabular}{lcl}
$\Delta \al _{\rm hadrons}^{(5)}(\mz)$   &$=$& $0.02768 \pm
0.00036$~\cite{BP03}\\ &&$0.02769 \pm 0.00018$~\cite{HMNT03}
\end{tabular}
(see also \cite{Pietrzyk,HMNT03}). Other aspects concerning
$\alpha(M_Z)$ were discussed in~\cite{Trentadue,Vicini} (see also~\cite{JF85}).

\section{EVALUATION OF $a_\mu \equiv (g-2)_\mu/2$}

The leading non-perturbative hadronic contribution to 
$\amuh$, given by the diagram\\[-4mm]

\centerline{%
\includegraphics[scale=0.75]{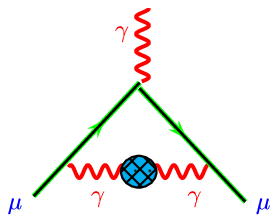}}
\noi can be obtained again in terms of
$R_\gamma(s)$ via the dispersion integral \small
\ba
\amuh &=& \left(\frac{\alpha m_\mu}{3\pi}
\right)^2 \bigg(\;\;\;
\int\limits_{4 m_\pi^2}^{E^2_{\rm cut}}ds\,
\frac{ {R^{\mathrm{data}}_\gamma(s)}\;\hat{K}(s)}{s^2}
\nn \\ &&~~~+ \int\limits_{E^2_{\rm cut}}^{\infty}ds\,
\frac{{R^{\mathrm{pQCD}}_\gamma(s)}\;\hat{K}(s)}{s^2}\,\,
\bigg)
\ea \normalsize
where $\hat{K}(s) \in [0.63\cdots,1.0]$ is a bounded monotonically
increasing function.

\begin{figure}[ht]
\begin{picture}(120,60)(-20,30)
\includegraphics[scale=0.45]{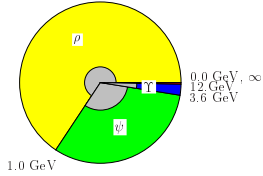}
\end{picture}

\caption{The distribution of contributions and errors
(shaded areas scaled up by 10) for $\amuh\;$. }
\label{fig:gmusta}

\vspace{-7mm}

\end{figure}

Again the experimental error of $R$ implies the main theoretical
uncertainty in the prediction of $\amu$. Since the low energy
contribution is enhanced by $1/s^2$, about $\sim 67\% $ of the error
on $\amuh$ comes from the low energy region $4m_\pi^2<m^2_{\pi
\pi}<M_\Phi^2$. This is why the $\rho$ resonance region dominated by the
$\epm \to \pi^+\pi^-$ channel plays a prominent role in a precise
prediction of $\amuh$. My $\epm$--data based update yields (see
Fig.~\ref{fig:gmusta})\footnote{Other recent $\epm$--data based
results in units of $10^{-10}$ are $684.7 \pm 7.0$~\cite{DEHZ} and
$683.1 \pm 6.2$~\cite{HMNT03}.  The $\tau$--data based result
is~\cite{DEHZ} $\amu^{\mathrm{had}(1)}=(701.9 \pm
6.1)\:\times\:10^{-10}$.}\\[-5mm]
\be
\amu^{\mathrm{had}(1)}=(694.8 \pm 8.6)\:\times\:10^{-10} \ep
\ee
With this estimate I get
\ba
\amu^\mathrm{the} &=& (11\:659\:179.4  \pm 8.6_{\rm had } \pm 3.5_{\rm LBL} 
\nn \\ && \pm 0.4_{\mathrm{QED+EW}}) \times 10^{-10}
\ea
which compares to the most recent experimental result~\cite{BNL,BLR}
\be
\amu^\mathrm{exp} = 11\:659\:203 \pm 8) \times 10^{-10} \ep
\ee
The ``discrepancy'' $|\amu^\mathrm{the} - \amu^\mathrm{exp}| = (23.6
\pm 12.3) \times 10^{-10}$ corresponds to a deviation of about 1.9
$\sigma$. For other recent estimates
see~\cite{DEHZ,Davier3,HMNT03,Colangelo}. The status of the theory has
been reviewed in~\cite{Nyffeler}. For a review of the history of
$\amu$ experiments see~\cite{Farley}.

\vspace{-10mm}

Some comments: What we need for inserting into the dispersion
integrals is on the one hand the one particle irreducible (1pi)
``blob'', which corresponds to the VP undressed cross-section (pion
form factor)\\[-4mm]

\centerline{%
\includegraphics[scale=0.6]{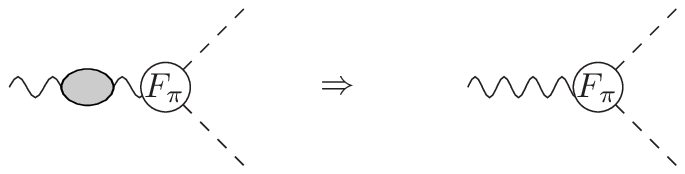}}

\vspace*{-4mm}

\be
\sigma_{\pi\pi}^{(0)} (s)=\sigma_{\pi\pi} (s)\:(\alpha/\alpha(s))^2 \ep
\ee
On the other hand, the hadronic 1pi ``blob'' should include 
photonic corrections:
\centerline{%
\includegraphics[scale=0.6]{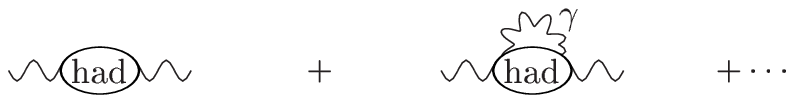}.}
Thus one has to add the theoretical prediction for FS radiation
(including full photon phase space):
\be
\sigma^{(\gamma)}_{\pi\pi}(s) = \sigma^{(0)}_{\pi\pi}(s)
\:\left(1+\eta(s)\frac{\alpha}{\pi}\right)
\ee
to order $O(\alpha)$, where $\eta(s)$, to the extent that scalar QED
(sQED) for the pion photon interactions is assumed as a model, is a
known correction factor (Schwinger 1989). The corresponding $O(\alpha)$
contribution to the anomalous magnetic moment of the muon is
\be
\delta^{\gamma} \amuh=(38.6 \pm 1.0) \times 10^{-11} \ep
\ee
In order to get a model independent result one has to measure 
FSR~\cite{Gluza:2002ui}, preferably in an energy scan experiment.

When including or excluding (note that only more or less hard real
photons can be cut away) final state photons (FSR) one has to be aware
of the Kinoshita, Lee, Nauenberg (KLN) theorem at work. The fully
inclusive cross-section, including virtual ($V$), soft ($S$) and all
hard ($H$) photons, is of the form $\sigma_0 \: (1 +
\frac{\alpha}{\pi}C)$~ $C$ a constant of $O(1)$ like 3/4 for
$\mu$--pair, 3 (in sQED) for $\pi$--pair production, i.e. typically
this correction is small (and positive) as a matter of cancellations
between $V+S$, on the one hand, and $H$, on the other hand. A typical
size of the FSR contribution is $\sim$ .2\%. Exclusive quantities, in
contrast, involve large logs on the photon cut energy as well as
collinear logs.  According to Bloch and Nordsieck, $V+S$ cannot be
separated (separately IR divergent) and there are potentially large
logs proportional to $\ln (\sqrt{s}/E_{\gamma {\rm cut}})$ and $\ln
(s/m_\pi^2)$ which multiply $\frac{\alpha}{\pi}$ and the complementary
hard photon part (integrated over full phase space) $H$ exhibits the
same logs as $V+S$ but precisely of opposite sign.  All logs cancel in
the sum (inclusive measurement). Consequences are the following: 1)
cutting out all hard photons and subtracting $V+S$ using sQED one
expects a .5\% model ambiguity (sQED vs. fQED) 2) adding the missing
$H$ part using the same model (as used for subtracting $V+S$) makes
the correction small (for any model) and thus also reduces the model
ambiguity to about .1\% (sQED vs. fQED)~\cite{Gluza:2002ui}. The true
model error of course cannot be obtained this way.

As mentioned before the low energy region plays a particularly
important role for the evaluation of $\amuh$. The region is dominated
by the pion form factor $F_\pi(s)$, for which severe theory
constraints exist. In fact {\em analyticity}, {\em unitarity} and the
{\em chiral limit} relate space--like data, $\pi \pi$--scattering
phase shifts and time--like data in a stringent way
(Omn\`es--Muskhelishvili approach). Work in progress will allow us to
get a substantially improved low energy
contribution~\cite{Leutwyler:2002hm,Colangelo,Portoles}. For earlier attempts in this
direction see~\cite{Casas:yw,Geshkenbein:1998gu}.

\section{$e^+e^-$--CROSS SECTIONS VIA \hfill \\ 
$\tau$--DECAY SPECTRAL FUNCTIONS} To the extent that the charged
vector current is (approximately) conserved (CVC), the iso-vector part
of $\sigma(e^+e^- \to {\rm hadrons})$ may be obtained by an iso-spin
rotation from $\tau$--decay spectra. The precise relation derives from
the diagrams
\centerline{%
\includegraphics[scale=0.6]{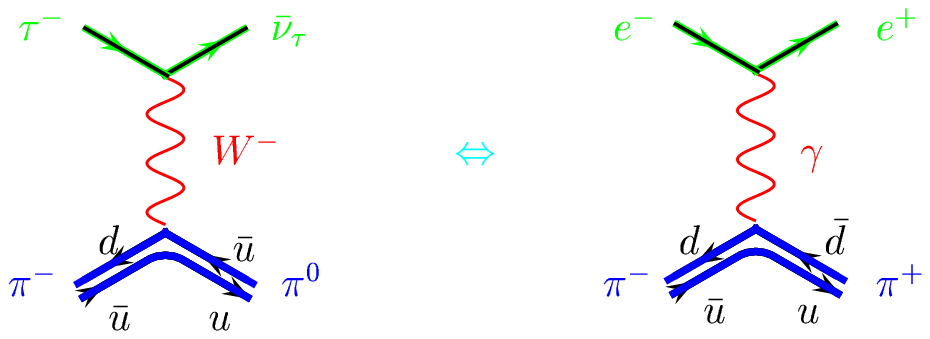}}
$$ \tau^- \ra X^- \nu_\tau \;\;\;
\leftrightarrow \;\;\; e^+ e¯ \ra
X^0$$ where $X^-$ and $X^0$ are hadronic states related by an
iso-spin rotation. The $\epm$ cross--section is then given by
\be
\sigma_{\epm \ra X^0}^{I=1}= \frac{4 \pi
\al^2}{s}v_{1,X^-}\;\;,\;\;\;
\sqrt{s} \leq M_\tau
\ee
in terms of the $\tau$ spectral function $v_1$. In principle, the such
enhanced ``$\epm$ data set'' is able to improve substantially the
knowledge of the $\pi^+\pi^-$ channel ($\rho$--resonance contribution)
which is the dominating contribution (72\%) to $\amuh$. Note that the
$\tau$--data can replace $\epm$--data only partially, because the
$I=1$ part only accounts for about 75\% of the total, summed over all
channels. The inclusion of the $\tau$ data (from ALEPH) has been
pioneered in Ref.~\cite{ADH98} at a time when errors of the CMD-2 data
were a factor of two larger and there was no reason not to combine the
$\epm$ and $\tau$ data. In particular for $\amuh$ the reduction of
the error was about 30\%:
$\delta a_\mu ~:~ 15.6\times 10^{-10} \ra  10.2 \times 10^{-10}$
while $\delta \Delta \alpha ~:~ 0.00067  \ra  0.00065 \ep$

\begin{figure}[htb]

\centerline{%
\includegraphics[scale=0.55]{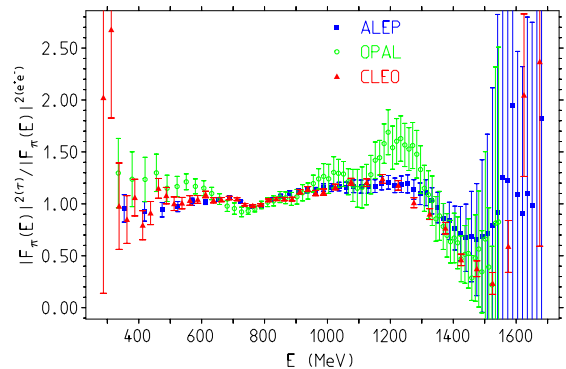}}

\vspace{-10mm}

\caption{Comparison of the $\tau$-data from ALEPH, CLEO and OPAL,
normalized to the CMD-2 fit. Note that the agreement between
ALEPH/CLEO and OPAL is not very good.}
\label{fig:taucomp}

\vspace{-7mm}

\end{figure}

In the meantime, CMD-2 was able to substantially reduce the systematic
error in $\epm \to \pi^+\pi^-$ and OPAL and CLEO came up with
independent measurements of $\tau$ spectral functions.  Obviously,
before data can be combined all kind of iso-spin breaking effects have
to be taken into account. Taking into account the established iso-spin
breaking effects~\cite{CEN} a comparison of the $\tau$-data is shown
in Fig.~\ref{fig:taucomp}. As already mentioned, the reanalysis
based on the most recent data revealed a substantial discrepancy
between the $\epm$ and the $\tau$ data to an extent which made a
combination of the data impossible~\cite{DEHZ}.


\begin{figure}[htb]
\centerline{%
\includegraphics[scale=0.55]{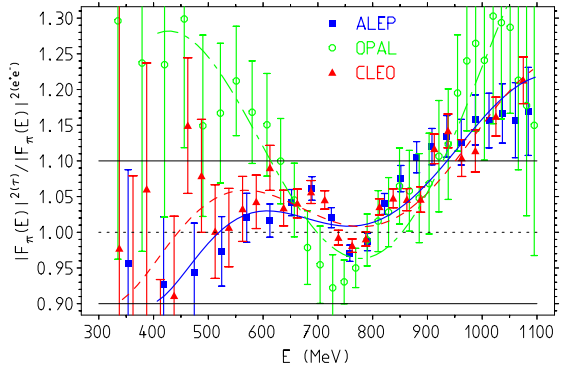}}

\vspace{-10mm}

\caption{The ratio between $\tau$--data sets from ALEPH, OPAL and CLEO 
and the $I=1$ part of the CMD-2 fit of the $\epm$--data. The curves
which should guide the eye are fits of the ratios using 8th order
Tschebycheff polynomials.}
\label{fig:ratio}

\vspace{-7mm}

\end{figure}

What may be the origin of the large discrepancy? This question has
been analyzed recently in~\cite{Ghozzi:2003yn}. For comparing $\epm$
and $\tau$ data we may use a Gounaris-Sakurai (GS) parametrization of the
pion form-factor. It is crucial to compare ``commensurate'' data: 

{\bf 1)} for $\epm$ subtract VP, subtract FSR, undo
$\rho-\omega$-mixing and fit for $m_{\rho^0}$ and $\Gamma_{\rho^0}$ at
fixed background (i.e. all other parameters of the GS parametrization
held fixed) which yields~\cite{Ghozzi:2003yn}:
$$m_{\rho^0}=773.1\pm0.6~~,~~~\Gamma_{\rho^0}=146.4 \pm 1.0$$
{\bf 2)} apply the iso-spin corrections~\cite{CEN,Erler:2002mv} 
$$r_{\mathrm{IB}}(s)=\frac{1}{G_{\rm EM}(s)} \:
\frac{\beta^3_{\pi^-\pi^+}}{\beta^3_{\pi^- \pi^0}} 
\:\frac{S_{\mathrm{EW(old)}}}{S_{\mathrm{EW(new)}}}
$$ to the $\tau$ data and fit the corrected $\tau$ data for
$m_{\rho^-}$ and $\Gamma_{\rho^-}$ at fixed background, which yields
$$ m_{\rho^-}=775.8\pm0.7~~,~~~\Gamma_{\rho^-}=147.8 \pm 1.0 \ep $$
Thus the discrepancy shown in Fig.~\ref{fig:ratio} may be interpreted
as a shift in the mass and the width\footnote{We have $\Delta
m_\rho=2.7 \pm 0.9 [3.1 \pm 0.9]$ MeV; $\Delta \Gamma_\rho =1.4 \pm
1.4 [1.8 \pm 1.6]$ MeV, in brackets the values from~\cite{Davier1}.} of
the $\rho$ as the leading effect~\cite{Ghozzi:2003yn} (see
also~\cite{Davier1}).


Since both data set may be fitted rather well by a GS-formula, it is
not very surprising that one can easily fit the ratio to unity within
errors by letting $m_{\rho^-}$ and $\Gamma_{\rho^-}$ float in the
numerator. The fit yields back $m_{\rho^0}$ and $\Gamma_{\rho^0}$;
this is not completely trivial as it works over the whole range
shown. Appraisal of the result:

\myit We now have two sets of reasonably consistent data: 
ALEPH and CLEO $\tau$--data vs. CMD-2 and KLOE $\epm$--data.

\myit What effect is able to to give 10\% in a resonance tail? Answer:
a shift of the energy by about 1\%. This could be a problem of energy
calibration, but this is very unlikely. Rather the physical resonance
parameters have no reason to be identical!

\myit In spite of possible experimental problems, there is no reason
to expect the neutral channel parameters to be the same as the charged
current ones. Example, pions: $m_{\pi^\pm}-m_{\pi^0}=4.5935\pm0.0005 \
\mv $. Why should this be very different for the $\rho$ a similar (same
quark content) bound state? I do not think that the Goldstone-boson
nature of the pions necessarily makes such an iso-spin breaking very
different.

Our conclusion: 

\myit the observed discrepancy is a so far unaccounted
iso-spin breaking effect, which the $\tau$-data have to be corrected for!

\myit Relevant for calculating $\amuh$ are the $\epm$--data in first place.

\myit How to include the $\tau$--data then? One should know 
$\Delta m_{\rho}$ and $\Delta \Gamma_{\rho}$ from elsewhere. Other
parameters like $m_{\rho'}$ and $\Gamma_{\rho'}$ are affected as well,
but his is a higher order effect.

\myit In spite of the fact that the peak has to be shifted downwards,
$\amuh$ does not increase as naively expected, in view of the $s^{-2}$
weight in the integral. As it should be, it rather decreases, because
the width also substantially goes down and over-compensates the effect
of the mass shift.

\myit If you don't like the idea: notice that the $\frac{\partial \amuh}{
\partial m_\rho [\Gamma_\rho]}$ is very large so that this quantity is 
very sensitive to the parameters. If it is not a true iso-spin breaking
effect, it would have to be an energy calibration problem.

My conclusion: we are back to \underline{one} prediction for
$a_\mu$  at $2\sigma$ from the experimental value!

\section{CONTROLLING PQCD VIA \hfill \\ THE ADLER FUNCTION}
\begin{figure}[htb]

%
\centerline{%
\includegraphics[scale=0.6]{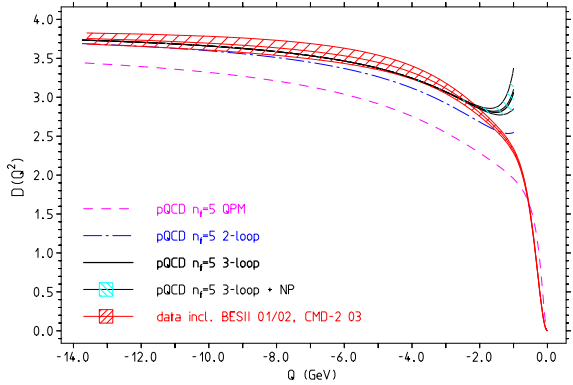}}

\vspace*{-7mm}

\caption{``Experimental'' Adler--function versus theory (pQCD + NP).}
\label{fig:adler}

\vspace*{-7mm}

\end{figure}

In particular for $\dalh$ the standard evaluation based on data is
very sensitive to errors of the data up to the $\Upsilon$ energies.
For future needs in precision physics, new experiments are needed which
measure $\sigha$ at the 1\% level up to 10 GeV. Is it possible to
depend less on inaccurate data and to use perturbative QCD instead?
The so called ``theory driven''
approach~\cite{Martin:1994we,DH98a,KS98,GKNS98,Erler98,HMNT03} relies
on replacing to a large extent the measured $R(s)$ by its
perturbatively calculated version (see~\cite{Steinhauser}). The quality
of such a procedure is hard to control beyond some perturbative
windows, because of non--perturbative effects present in $R(s)$
(resonances, thresholds). An alternative approach, which allows us to
control the applicability of pQCD in a much better way was proposed
long time ago by Adler: the Euclidean (space-like) method via the
Adler function
\small
\ba
D(-s) &\doteq&
\frac{3\pi}{\alpha}\:s\:\frac{d}{ds} \Delta \alpha_{\mathrm{had}}(s)
= -\left( 12 \pi^2 \right)\:s\: \frac{d\Pi'_{\gamma}(s)}{ds}\nn \\
D(Q^2)&=& Q^2\;\int\limits_{4 m_\pi^2}^{\infty}ds\,
\frac{R(s)}{\left( s+Q^2 \right)^2} \ep
\ea
\normalsize
It is an absolutely smooth
function which can be easily calculated from the experimental data on
the one hand and by pQCD on the other hand.
%
%
In Fig.~\ref{fig:adler} the present status is shown.

We see that in the space-like approach pQCD works well for $Q^2=-q^2 >
2.5$ GeV. There we may replace the experimental Adler function by the
pQCD one\footnote{Note that in the time-like approach pQCD works well only in
the ``perturbative windows'' 3.00 - 3.73 GeV, 5.00 - 10.52 GeV and 11.50 -
$\infty$~\cite{Steinhauser}.}.

In this approach we thus may calculate
\be
\Delta \alpha_{\mathrm{had}}(-Q^2) \sim \frac{\alpha}{3\pi} \int
dQ^{'2} \frac{D(Q^{'2})}{Q^{'2}}
\ee
\small
\bea
\Delta\alpha^{(5)}_{\rm had}(-M_Z^2)=
\left[\Delta\alpha^{(5)}_{\rm had}(-M_Z^2) -\Delta\alpha^{(5)}_{\rm
had}(-s_0)\right]^{\mathrm{pQCD}} \\
~~~~~~~~ +\:\Delta\alpha^{(5)}_{\rm had}(-s_0)^{\mathrm{data}}
\eea
\normalsize
and obtain, for $s_0=(2.5\, \gv)^2$: \small
\bea
\Delta\alpha^{(5)}_{\rm had}(-s_0)^{\mathrm{data}} = 0.007417 \pm
0.000086
\eea
\bea
\Delta\alpha^{(5)}_{\rm
had}(-M_Z^2) = 0.027613 \pm 0.000086 \pm 0.000149 \ep
\eea 
\normalsize
The second error comes from the variation of the pQCD parameters.
The largest uncertainty is due to the poor knowledge of the charm
mass.  I have taken errors to be 100\% correlated here. 

Link between space--like and time-like region is the difference
\small
$$\dahz-\Delta\alpha^{(5)}_{\rm had}(-M_Z^2)
=0.000038 \pm 0.000005 $$ 
\normalsize
which can be calculated in pQCD. Altogether we obtain the value given
earlier in (\ref{eq:adler}). For more details I refer
to~\cite{EJKV98,Jegerlehner:2003ip}.

The result clearly demonstrates how important the precise knowledge of
the QCD parameters is. In the Adler function approach the error is
dominated by the parameter uncertainties in the pQCD part (see
Fig.~{\ref{fig:aldist}}). Recent progress achieved by pQCD/SR (sum
rules, moment method) and lattice QCD (LQCD) is summarized in the
following Table~\cite{Steinhauser,Hoang,Rolf}:
\begin{table*}[hbt]
\begin{center}
\begin{tabular}{c|ccccc}
\hline
\hline
Ref & $\alpha_s(M_Z)$ & $\Lambda_{\overline{\rm
MS}}^{N_f=0}$ [MeV] &$m_s(m_s)$ [MeV] & $m_c(m_c)$ [GeV] & $m_b(m_b)$ [GeV] \\
\hline
PDG & 0.118(3) & - & 80-155 & 1.0-1.4 & 4.0-4.5 \\ 
\cite{Steinhauser}  & $0.124^{+0.011}_{-0.014}$& -  & - & 1.304(27) & 4.191(51) \\
\cite{Hoang}  & -  & - & - & 1.3(1) & 4.2(1) \\
\cite{Rolf} & - & 238(19)[Q]  & 97(4)[Q]   & 1.301(34)[Q] & 4.12(7)(4)[Q]\\  
\hline
\end{tabular}
\caption{New results for QCD parameters. [Q] quenched results from LQCD}
\label{tab:dal}
\end{center}
\end{table*}
In LQCD in not too far future full QCD results [unquenched]
extrapolated to the continuum limit will be available.

\vspace*{-2mm}

\section{STATUS AND OUTLOOK}
Conclusions: in addition to planned upgrades which have been reported,
a $\tau$--charm factory would be needed which should be able to
perform an energy scan between 2 and 3.6 GeV with 1\% accuracy. This
would help to satisfy requirements of future precision experiments
g-2, GigaZ, etc.

Last but not least: we need further progress in theory

\myit 
more careful study of iso-spin breaking in $\tau$--data vs. $\epm$--data

\myit  constraints to $F_\pi(s)$ from $\chi$PT,
unitarity and analyticity below the about $2M_K$.

\myit  still a theoretical challenge:
the hadronic light--by--light scattering contribution !

\myit 
Concerning the $\tau$--data vs. $\epm$--data discrepancy: we need more
careful check of radiative corrections (FSR, in particular) which have
been applied !

\myit 
Further progress in radiative corrections calculations is
needed for the processes involved in $R$--measurements

\myit
Also further progress in determination of the QCD parameters is
indispensable.

The big experimental challenge: one has to attempt cross-section
measurements at the 1\% level up to $J/\psi$[or even up to
$\Upsilon$].\\

{\bf Acknowledgments\\[-3mm]~}

I thank the organizers of the workshop SIGHAD03 for the kind
invitation and hospitality at Pisa. Thanks also to Mikhail
Kalmykov for helpful discussions and for reading the manuscript.



\begin{thebibliography}{99}
\frenchspacing

\bibitem{LEP} 
J.~Alcaraz {\it et al.} [LEP Collaborations], LEPEWWG/2003-01;
D.~Abbaneo {\it et al.} [LEP Collaborations], LEPEWWG/2002-02, 
hep-ex/0212036.

\bibitem{Zeuthen2003} Proc. of the Mini-Workshop on 
``Electroweak Precision Data And The Higgs Mass'', Zeuthen 2003, Eds.
S.~Dittmaier and K.~M\"onig,
K.~M\"onig,
hep-ph/0308133 and references therein.

\bibitem{Pietrzyk} B. Pietrzyk, ``Fit of S.M., results on $\alpha_{\rm
em}(M_Z)$ and influence on the Higgs mass'', these proceedings.

\bibitem{CMD2}
R.~R.~Akhmetshin {\it et al.}  [CMD-2 Collaboration],
hep-ex/0308008.

\bibitem{Fedotovitch} G. Fedotovitch, ``Precise measurements of hadronic 
cross section at VEPP-2M'', these proceedings.
 
\bibitem{CMD}
R.~R.~Akhmetshin {\it et al.}  [CMD-2 Collaboration],
Phys.\ Lett.\ B {\bf 527} (2002) 161.

\bibitem{BES}
J.~Z.~Bai {\it et al.}  [BES Collaboration],
Phys.\ Rev.\ Lett.\  {\bf 84} (2000) 594;
Phys.\ Rev.\ Lett.\  {\bf 88} (2002) 101802.

\bibitem{Hu} H. Hu, ``Status and future plans of $R$ measurement at BES'',
these proceedings.

\bibitem{ALEPH}
R.~Barate {\it et al.}  [ALEPH Collaboration],
Z.\ Phys.\ C {\bf 76} (1997) 15;
Eur.\ Phys.\ J.\ C {\bf 4} (1998) 409.

\bibitem{OPAL}
K.~Ackerstaff {\it et al.}  [OPAL Collaboration],
Eur.\ Phys.\ J.\ C {\bf 7} (1999) 571.

\bibitem{CLEO}
S.~Anderson {\it et al.}  [CLEO Collaboration],
Phys.\ Rev.\ D {\bf 61} (2000) 112002.

\bibitem{Davier1} M. Davier, ``$\tau$ spectral functions and QCD'',  
these proceedings.

\bibitem{tsai}
Y.~S.~Tsai,
Phys.\ Rev.\ D {\bf 4} (1971) 2821
[Erratum-ibid.\ D {\bf 13} (1976) 771].

\bibitem{ADH98} 
R.~Alemany, M.~Davier and A.~H\"ocker,
Eur.\ Phys.\ J.\ C {\bf 2} (1998) 123.

\bibitem{Hayashii} H. Hayashii, ``$\tau \to 2 \pi$ decay mode from Belle'',
these proceedings.  

\bibitem{DEHZ}
M.~Davier, S.~Eidelman, A.~H\"ocker and Z.~Zhang,
Eur.\ Phys.\ J.\ C {\bf 27} (2003) 497;
hep-ph/0308213.
  
\bibitem{CEN}
V.~Cirigliano, G.~Ecker and H.~Neufeld,
Phys.\ Lett.\ B {\bf 513} (2001) 361; JHEP {\bf 0208} (2002) 002.

\bibitem{Ecker} G. Ecker, ``Isospin violating and radiative corrections 
to $\tau \to 2  \pi$ decay'',
these proceedings.   

\bibitem{Eidelman03}
S.~Eidelman, private communication.

\bibitem{Narison:2003ur}
S.~Narison,
Phys.\ Lett.\ B {\bf 568} (2003) 231.

\bibitem{Dubnicka} S. Dubnicka, 
these proceedings.

\bibitem{RR} S.~Binner, J.~H.~K\"uhn and K.~Melnikov,
Phys.\ Lett.\ B {\bf 459} (1999) 279;
H.~Czy\.z and J.~H.~K\"uhn,
{\em Eur.\ Phys.\ J.} {\bf C18} (2001) 497;
G.~Rodrigo, H.~Czy\.z, J.~H.~K\"uhn and M.~Szopa,
Eur.\ Phys.\ J.\ C {\bf 24} (2002) 71;
J.~H.~K\"uhn and G.~Rodrigo,
Eur.\ Phys.\ J.\ C {\bf 25} (2002) 215;
H.~Czy\.z, A.~Grzelinska, J.~H.~K\"uhn and G.~Rodrigo,
Eur.\ Phys.\ J.\ C {\bf 27} (2003) 563;
hep-ph/0308312.

\bibitem{Khoze:2002ix}
V.~A.~Khoze {\it et al.},
Eur.\ Phys.\ J.\ C {\bf 25} (2002) 199.

\bibitem{KLOE}
A.~Aloisio {\it et al.}  [KLOE Collaboration],
hep-ex/0307051;
A.~Denig [KLOE Collaboration],
hep-ex/0311012.

\bibitem{Valeriani} B. Valeriani, ``Results on $R$ from KLOE'',
these proceedings.

\bibitem{Davier2} M. Davier, ``Results on $R$ from BABAR'',
these proceedings.
     
\bibitem{Eidelman2}  S. Eidelman, ``First results on $R$ from BELLE'',
these proceedings.

\bibitem{Dytman} S. Dytman, ``Results and plans on $R$ from CLEO'',
these proceedings.


\bibitem{Eidelman1} S. Eidelman, ``Plans on $R$ with VEPP-2000'',
these proceedings.

\bibitem{Denig} A. Denig, ``Future plans for Dafne'',
these proceedings.

\bibitem{Nyffeler}
A.~Nyffeler,
hep-ph/0305135;
``Theory of $(g-2)_\mu$'',
these proceedings.

\bibitem{Knecht:2001qf}
M.~Knecht and A.~Nyffeler,
Phys.\ Rev.\ D {\bf 65} (2002) 073034;
M.~Knecht, A.~Nyffeler, M.~Perrottet and E.~De Rafael,
Phys.\ Rev.\ Lett.\  {\bf 88} (2002) 071802.

\bibitem{Bijnens:1995xf}
J.~Bijnens, E.~Pallante and J.~Prades,
Nucl.\ Phys.\ B {\bf 474} (1996) 379;
Nucl.\ Phys.\ B {\bf 626} (2002) 410.

\bibitem{Hayakawa:1997rq}
M.~Hayakawa and T.~Kinoshita,
Phys.\ Rev.\ D {\bf 57} (1998) 465
[Erratum-ibid.\ D {\bf 66} (2002) 019902].

\bibitem{Hoefer:2001mx}
A.~Hoefer, J.~Gluza and F.~Jegerlehner,
Eur.\ Phys.\ J.\ C {\bf 24} (2002) 51.

\bibitem{Arbuzov:2003xf}
A.~B.~Arbuzov {\it et al.},
hep-ph/0308292.

\bibitem{Rodrigo}
G. Rodrigo, ``Radiative return at meson factories'',
these proceedings.

\bibitem{EJ95}
S.~Eidelman and F.~Jegerlehner,
Z.\ Phys.\ C {\bf 67} (1995) 585; see also:
F.~Jegerlehner,
Z.\ Phys.\ C {\bf 32} (1986) 195;
Nucl.\ Phys.\ Proc.\ Suppl.\  {\bf 51C} (1996) 131;
J.\ Phys.\ G {\bf 29} (2003) 101.

\bibitem{Czyz:2003gb}
H.~Czy\.z and E.~Nowak,
hep-ph/0310335.

\bibitem{Sibidanov} A. Sibidanov, ``Radiative corrections at CMD-2'',
these proceedings.

\bibitem{CaCa} C.M. Carloni Calame, ``The BABAYAGA event generator'',
these proceedings.

\bibitem{Scimemi} I. Scimemi,  ``E.M. in hadronic processes'',
these proceedings.

\bibitem{Kuhn} J. K\"uhn, ``Status and prospects of MC tools for ISR'',
these proceedings.

\bibitem{Zhao} Z. Zhao,  ``Status and prospects of $R$ measurements'',  
these proceedings.

\bibitem{BP03}
H.~Burkhardt and B.~Pietrzyk, Phys.\ Lett.\ B {\bf 513} (2001) 46.

\bibitem{HMNT03}
K.~Hagiwara {\it et al.}, 
Phys.\ Lett.\ B {\bf 557} (2003) 69;
T. Teubner, ``Hadronic contributions to $a_\mu$'',
these proceedings.

\bibitem{Trentadue} L. Trentadue,  ``The running of $\alpha_{\rm QED}$ in 
small angle Bhabha scattering'', these proceedings.

\bibitem{Vicini}  
G.~Degrassi and A.~Vicini,
hep-ph/0307122;
A. Vicini, ``Two-loop renormalization of the electric 
charge in the Standard Model'', these proceedings.

\bibitem{JF85}
F.~Jegerlehner and J.~Fleischer,
Phys.\ Lett.\ B {\bf 151} (1985) 65;
Acta Phys.\ Polon.\ B {\bf 17} (1986) 709;
F.~Jegerlehner, hep-ph/0105283.

\bibitem{BNL} 
G.~W.~Bennett {\it et al.}  [Muon g-2 Collaboration],
Phys.\ Rev.\ Lett.\  {\bf 89} (2002) 101804
[Erratum-ibid.\  {\bf 89} (2002) 129903].

\bibitem{BLR}  B. Lee Roberts, ``Results and prospects on the measurement of 
$(g-2)_\mu$'', these proceedings.

\bibitem{Davier3} M. Davier, ``Hadronic contributions to $a_\mu$'',
these proceedings.

\bibitem{Colangelo} G. Colangelo, ``Hadronic contributions to $a_\mu$'',  
these proceedings.
 
\bibitem{Farley} F. Farley,  ``46 years of $g-2$'',
these proceedings.

\bibitem{Gluza:2002ui}
J.~Gluza {\it et al.}, 
Eur.\ Phys.\ J.\ C {\bf 28} (2003) 261.

\bibitem{Leutwyler:2002hm}
H.~Leutwyler,
hep-ph/0212324.
 
\bibitem{Portoles} J. Portoles, ``The hadronic cross-section in the 
resonance energy region'',
these proceedings.  

\bibitem{Casas:yw}
J.~A.~Casas, C.~Lopez and F.~J.~Yndurain,
Phys.\ Rev.\ D {\bf 32} (1985) 736;
J.~F.~De Troconiz and F.~J.~Yndurain,
Phys.\ Rev.\ D {\bf 65} (2002) 093001.

\bibitem{Geshkenbein:1998gu}
B.~V.~Geshkenbein,
Phys.\ Rev.\ D {\bf 61} (2000) 033009.

\bibitem{Ghozzi:2003yn}
S.~Ghozzi and F.~Jegerlehner,\\
hep-ph/0310181.

\bibitem{Erler:2002mv}
J.~Erler,
hep-ph/0211345.

\bibitem{Martin:1994we}
A.~D.~Martin and D.~Zeppenfeld,
Phys.\ Lett.\ B {\bf 345} (1995) 558.

\bibitem{DH98a} M. Davier, A. H\"ocker, \PLB {\bf 419}, 419 (1998); ibid.
\PLB {\bf 435}, 427 (1998).

\bibitem{KS98} J.H. K\"uhn, M. Steinhauser, \PLB {\bf 437}, 425 (1998)

\bibitem{GKNS98} S. Groote, J.G. K\"orner, N.F. Nasrallah, K. Schilcher,
                 \PLB {\bf 440}, 375 (1998)

\bibitem{Erler98} J. Erler, hep-ph/9803453

\bibitem{Steinhauser}
J.~H.~K\"uhn and M.~Steinhauser,
Nucl.\ Phys.\ B {\bf 619} (2001) 588
[Erratum-ibid.\ B {\bf 640} (2002) 415];
M. Steinhauser, ``Impact of $\sigma(e^+e^- \to
{\rm hadrons})$ measurements on the parameters of the S. M.'',
these proceedings.

\bibitem{EJKV98}
S.~Eidelman, F.~Jegerlehner, A.~L.~Kataev and O.~Veretin,
Phys.\ Lett.\ B {\bf 454} (1999) 369;\\
F. Jegerlehner, 
hep-ph/9901386.

\bibitem{Jegerlehner:2003ip}
F.~Jegerlehner,
hep-ph/0308117.

\bibitem{Hoang} A. Hoang,  ``Status of quark masses'', 
these proceedings.

\bibitem{Rolf} 
M.~Della Morte {\it et al.} [ALPHA collaboration],
hep-lat/0209023;\\
J.~Rolf and S.~Sint  [ALPHA Collaboration],
JHEP {\bf 0212} (2002) 007;
J. Rolf,  ``Strong coupling and quark masses from 
lattice QCD'',  these proceedings.

\end{thebibliography}
\end{document}